\documentclass[sigconf]{acmart}

\AtBeginDocument{%
  }

\setcopyright{acmlicensed}
\copyrightyear{2025}
\acmYear{2025}
\acmDOI{XXXXXXX.XXXXXXX}

\acmConference[The Web Conference '25]{}{April 28 - May 2}{Sydney, Australia}
\acmISBN{978-1-4503-XXXX-X/18/06}

\usepackage{xcolor}
\usepackage[normalem]{ulem}

\usepackage{amsmath}
\usepackage{algorithm}
\usepackage{algorithmic}
\usepackage{tikz}
\usepackage{pgfplots}
\pgfplotsset{compat=1.16}
\usepackage{xcolor}
\usepackage{float}
\usepackage{graphicx}
\usepackage{multirow}
\usepackage{adjustbox}
\usepackage{caption}
\usepackage{multirow}
\usepackage{adjustbox}
\usepackage{caption}
\usepackage{booktabs}
\usepackage{xcolor}
\usepackage{caption}
\usepackage{subcaption}
\usepackage{stfloats} 
\usepackage{placeins} 
\usepackage{tikz}
\usepackage{pifont}   
\usetikzlibrary{positioning,shadows}
\usepackage{xcolor} 
\usepackage{fontawesome}

\usetikzlibrary{shapes.geometric, arrows.meta, positioning}

\tikzstyle{block} = [rectangle, rounded corners, minimum width=3cm, minimum height=2cm, text centered, draw=gray, fill=gray!20]

\tikzstyle{block_p2p} = [rectangle, rounded corners, minimum width=3cm, minimum height=4cm, text centered,  fill=blue!60,text=white, font=\small]
\tikzstyle{block_test_set} = [rectangle, rounded corners, minimum width=2cm, minimum height=4cm, text=white,  fill=green!60!black!60
,text centered, font=\small]
\tikzstyle{block_reddit} = [rectangle, rounded corners, minimum width=6cm, minimum height=8.5cm, text centered, fill=yellow, font=\small,text=white]
\tikzstyle{block_GAT_adv} = [rounded corners, minimum width=12cm, minimum height=4cm, draw=black, fill=gray!2, font=\small,line width=1]

\tikzstyle{line} = [thick, ->, >=stealth]
\tikzstyle{doubleline} = [thick, <->, >=stealth]
\tikzstyle{data} = [rectangle, rounded corners, minimum width=3cm, minimum height=3.5cm, text centered, draw=black, fill=gray!0, font=\small]
\tikzstyle{decision} = [diamond, minimum width=3cm, minimum height=1.5cm, text centered, draw=black, fill=gray!20, font=\small]
\tikzstyle{label} = [rectangle, rounded corners, minimum width=2.5cm, minimum height=1.2cm, text centered, draw=black, fill=gray!10, font=\small]

\begin{document}

\title{CrediRAG: Network-Augmented Credibility-Based Retrieval for Misinformation Detection in Reddit}


\author{Ashwin Ram}
\affiliation{%
  \institution{The University of Texas at Austin}
  \city{Austin}
  \state{Texas}
  \country{USA}}
\email{ashwin.ram@utexas.edu}

\author{Yigit Ege Bayiz}
\affiliation{%
 \institution{The University of Texas at Austin}
  \city{Austin}
  \state{Texas}
  \country{USA}}
\email{egebayiz@utexas.edu}

\author{Arash Amini}
\affiliation{%
 \institution{The University of Texas at Austin}
  \city{Austin}
  \state{Texas}
  \country{USA}}
\email{a.amini@utexas.edu}

\author{Mustafa Munir}
\affiliation{%
 \institution{The University of Texas at Austin}
  \city{Austin}
  \state{Texas}
  \country{USA}}
\email{mmunir@utexas.edu}

\author{Radu Marculescu}
\affiliation{%
 \institution{The University of Texas at Austin}
  \city{Austin}
  \state{Texas}
  \country{USA}}
\email{radum@utexas.edu}

\renewcommand{\shortauthors}{A. Ram et al.}

\begin{abstract}
Fake news threatens democracy and exacerbates the polarization and divisions in society; therefore, accurately detecting online misinformation is the foundation of addressing this issue. We present CrediRAG, the first fake news detection model that combines language models with access to a rich external political knowledge base with a dense social network to detect fake news across social media at scale.
CrediRAG uses a news retriever to initially assign a misinformation score to each post based on the source credibility of similar news articles to the post title content. CrediRAG then improves the initial retrieval estimations through a novel weighted post-to-post network connected based on shared commenters and weighted by the average stance of all shared commenters across every pair of posts. We achieve 11\% increase in the F1-score in detecting misinformative posts over state-of-the-art methods. Extensive experiments conducted on curated real-world Reddit data of over 200,000 posts demonstrate the superior performance of CrediRAG on existing baselines. Thus, our approach offers a more accurate and scalable solution to combat the spread of fake news across social media platforms.
\end{abstract}

\ccsdesc[500]{Information systems~Natural language processing}
\ccsdesc[500]{Information systems~Social networks}
\ccsdesc[500]{Information systems~Web Searching and Document Retrieval}
\ccsdesc[500]{Information systems~Language models}

\keywords{Fake News Detection, Retrieval-Augmented Generation, Social Networks, LLMs, Graph Attention Networks, Computational Social Science}

\maketitle

\section{Introduction}

\begin{figure*}[ht]
    \centering
    \definecolor{reddit_color}{rgb}{1, 0.337, 0}
\resizebox{\textwidth}{!}{%

\begin{tikzpicture}[node distance=2cm]
    \node (reddit) [block_reddit,xshift=-2cm, fill=reddit_color,font=\Large,text=white] {};
    \node (reddit_text)[xshift=-2cm,yshift=3.5cm, font=\LARGE, text=white] {\textbf{Reddit Posts}};    
    \node (white_reddit)[block_reddit,xshift=-2cm, yshift=-.65cm, fill=reddit_color,font=\Large,text=white,minimum height=7cm,
    minimum width=5.8cm] {};

\node[draw=green!70!black, fill=green!5, thick, rectangle, rounded corners, drop shadow, minimum width=4.5cm, text width=4.5cm, align=left, inner sep=8pt, below=0.4cm of reddit_text] (real) {
    \textbf{\textcolor{black}{\large Authentic Reddit Post}} \hfill \textcolor{green!60!black}{\Large \checkmark}\\
    \vspace{5pt}
    \textcolor{blue}{NASA} confirms that the James Webb Space Telescope has successfully captured its first images of distant galaxies.
};

\node[draw=red!70!black, fill=red!5, thick, rectangle, rounded corners, drop shadow, minimum width=4.5cm, text width=4.5cm, align=left, inner sep=8pt, below=0.4cm of real] (fake) {
    \textbf{\textcolor{black}{\large Fake Reddit Post}} \hfill \textcolor{red}{\Large \ding{55}}\\
    \vspace{5pt}
    \textcolor{blue}{The COVID-19 vaccine} contains microchips for government tracking.
};

\node[draw=none, fill=none, below=0.1cm of fake, align=center, text=white!80] (ellipsis) {
    \textbf{\huge \vdots}
};
    \node (P2P_1) [block_p2p, right of=reddit, xshift=3cm,yshift=2.25cm,text width=3cm,fill=gray!100,text=white] {Complete \textbf{Post2Post} Network\\(Algorithm \ref{alg:post2post}, Algorithm \ref{alg:weightedge})};
    \node (adversarial) [block_GAT_adv, right of=P2P_1, xshift=6cm,dashed]{};
    \coordinate[ right of =reddit,yshift=2.25cm,xshift=1cm] (aux_reddit_1);
    \coordinate[ right of =reddit,yshift=-1.25cm,xshift=1cm] (aux_reddit_2);
    \coordinate[ below of =adversarial,yshift=-1,xshift=1.5cm] (aux_adv_1);
    \node[ above of =adversarial,yshift=.2cm,xshift=-3.2cm] (aux_adv_2){\textbf{Adversarial Training for GAT}};
    \node (unknown) [block_test_set, below of=P2P_1, yshift=-1.5cm,text width=2cm,xshift=-.5cm,fill=reddit_color,minimum height=2cm] {Test Set Sampling};
    \node (p2p) [block_p2p, right of=unknown, xshift=1.cm,text width=3cm,fill=gray,minimum height=2cm] {Unlabeled \textbf{Post2Post} Network\\(Algorithm \ref{alg:post2post}, Algorithm \ref{alg:weightedge})};
    \node (rag) [block_p2p, right of=p2p, xshift=1.75cm,text width=3.5cm,fill=gray!90,minimum height=2cm] {Generating Initial\\ Labels By \textbf{RAG}};

    \node (trained) [block_test_set,text width=1cm,right of=rag, xshift=1.25cm,fill=gray!70,minimum height=2cm] {\textbf{Trained GAT}};
    \node (refined) [block_p2p, right of=trained, xshift=1cm,fill=gray!65,minimum height=2cm] {\textbf{Refined Labels}};


    \node (rag_fig)[block_GAT_adv,below of =rag,yshift=-.5cm,minimum height=2.5cm,minimum width =10,]{
    \includegraphics[width=10cm]{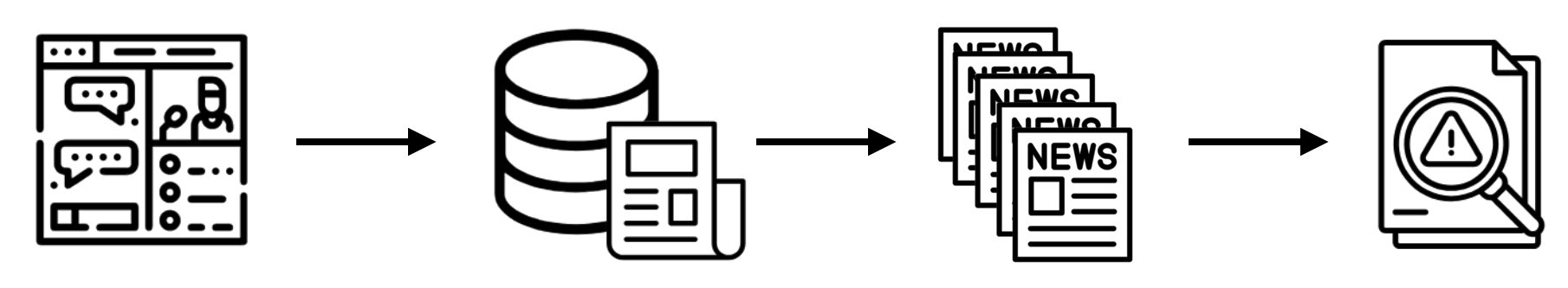}};
    \node (rag_fig1)[above of =rag_fig,xshift=-4.1cm,yshift=-1cm,font=\footnotesize]{Query};
    \node (rag_fig1)[above of =rag_fig,xshift=-1.4cm,yshift=-1cm,font=\footnotesize]{News Data Set};
    \node (rag_fig1)[above of =rag_fig,xshift=1.5cm,yshift=-1cm,font=\footnotesize]{Most Similar Articles};
    \node (rag_fig1)[above of =rag_fig,xshift=4.2cm,yshift=-1cm,font=\footnotesize]{Label};
    
    \node[above of=adversarial,draw=gray, fill=red!10, text width=11cm, rounded corners,line width=.5mm,font=\footnotesize,yshift=-.75cm,inner sep=5pt] (step1) 
    {\textbf{Step 1: Label Flipping (15\% of Labels)} \\
    Flip the labels of 15\% of the posts, turning fake labels into real and vice versa.};
    
    \node[below of=step1,draw=gray, fill=blue!10, text width=11cm, rounded corners,line width=.5mm,font=\footnotesize,yshift=.75cm,inner sep=5pt] (step2) 
    {\textbf{Step 2: Adversarial Training} \\
    Train the Graph Attention Network (GAT) to correct the flipped labels by challenging the model with the adversarial inputs.};
    
    \node[below of=step2,draw=gray, fill=yellow!10, text width=11cm, rounded corners,line width=.5mm,font=\footnotesize,yshift=.75cm,inner sep=5pt] (step3) 
    {\textbf{Step 3: GAT Refinement}\\
    The GAT corrects the labels and refines the model over successive iterations.};

    \draw[font=\large, fill=white] (0.4,3.7) circle (0.4);
    \node at (0.4,3.7) {\LARGE \textbf{A}};
    \draw[font=\large, fill=white] (4.0,3.7) circle (0.4);
    \node at (4.0,3.7) {\LARGE \textbf{B}};
    \draw[font=\large, fill=white] (16.45,3.7) circle (0.4);
    \node at (16.45,3.7) {\LARGE \textbf{C}};
    
    
    \draw [line] (aux_reddit_1) -- (P2P_1);
    \draw [line] (aux_reddit_2) -- (unknown);
    \draw [line] (P2P_1) -- (adversarial);
    \draw [line] (unknown) -- (p2p);

    \draw [line] (aux_adv_1) -- (trained);
    \draw [line] (rag) -- (trained);
    
    \draw [line] (trained) -- (refined);
    \draw [line] (p2p) -- (rag);
    
    \draw [line width=.5mm] (rag) -- (rag_fig);
    

\end{tikzpicture}}
    \caption{
    Overall Framework of \textit{CrediRAG}. The social media posts \textcircled{a} are the nodes in the weighted graph. We use Retrieval-augmented generation (RAG) to obtain all related news articles to a given post. The average credibility of all sources of retrieved articles is used to give an initial estimate of the misinformative level of every post in the graph. A corrective graph attention network (GAT) shown in \textcircled{c} is adversarially trained to refine labels based on the post-to-post network \textcircled{b}. This GAT corrects all of the RAG labels to give a final estimate of the binary label for each node.
    }
    \label{fig:methodology}
\end{figure*}
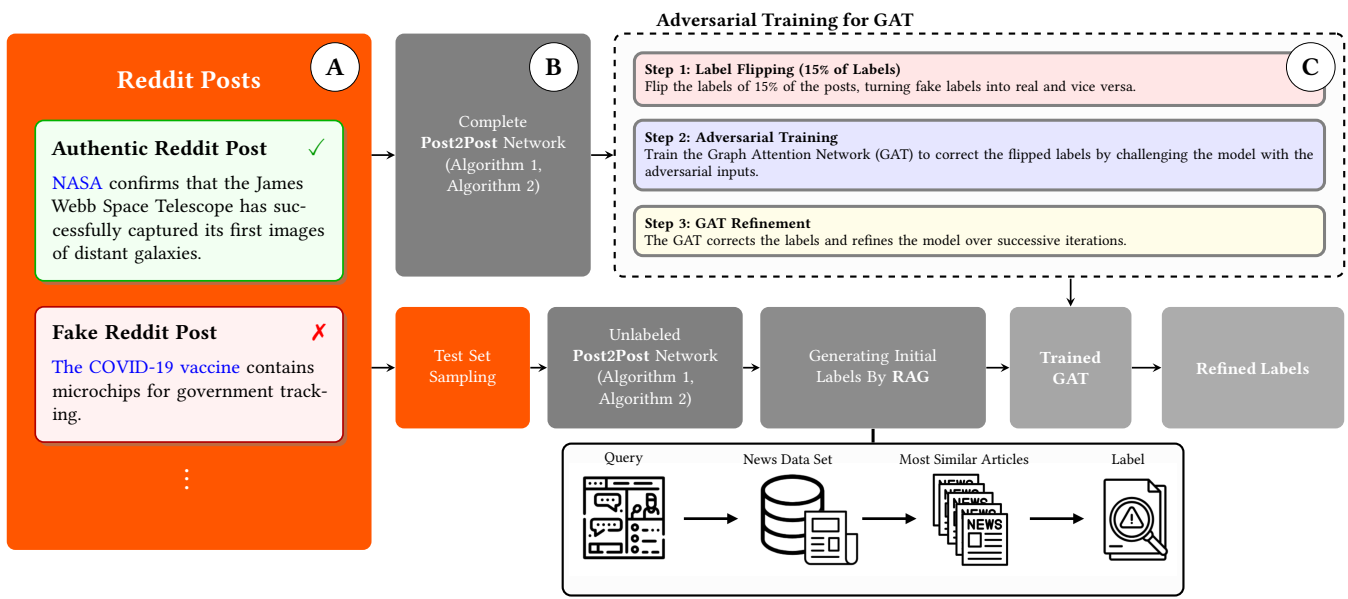

The increasing reliance on social media as a primary source of news consumption has dramatically reshaped the information landscape. Approximately two-thirds of American adults now access news through platforms like Facebook, Twitter, and Reddit, with Facebook being the most popular for news consumption \cite{Gottfried2016}. While social media has increased exposure to the news, particularly among younger and less-engaged audiences \cite{Fletcher2018}, it has also created a fertile ground for the spread of fake news and misinformation. Around $23\%$ of social media users occasionally share fake news, with over $60\%$ of users experiencing confusion about the veracity of news due to misinformation \cite{Nambiar2022}. Misinformation on social media is a critical issue because it has real-world consequences, particularly during crises like the COVID-19 pandemic, where it fueled vaccine hesitancy, led to harmful health practices, and increased mortality rates \cite{gisondi2022deadly, samia_tasnim_2019}.
Beyond public health, misinformation has caused widespread psychological distress and social tensions and eroded public trust in institutions, undermining the societal capacity to respond effectively to emergencies \cite{mendes2021impact, leng2021misinformation}. As such, automatically detecting misinformation became crucial because it can enable timely intervention to prevent the widespread harm caused by false information. 

Most social media platforms, including \textit{Reddit} and \textit{Twitter} have a structure where users can create posts and engage with them through comments and other forms of interaction. This hierarchical structure typically forms a tree-like pattern, with the original post at the root and subsequent comments and replies branching off. These interactions offer valuable network information, capturing the flow and propagation of content within the platform. Such interaction-driven engagement data has been utilized in various research areas, such as bot detection \cite{Hurtado2019}, where the interaction patterns between users and posts are analyzed to identify suspicious behavior. Reddit’s unique community-based model, where subreddits focus on specific topics and the interaction between posts can reveal news source credibility patterns, is rich for studying these dynamics \cite{Amini2024, bayiz2024susceptibility, sitaula2019credibilitybasedfakenewsdetection, leite2023detectingmisinformationllmpredictedcredibility, kozik2024explainablefakenewsdetection, 10.1145/3331184.3331285}. Importantly, this interaction-based approach is not limited to Reddit, but is generalizable to other social media platforms with similar comment chain structures, making it a versatile approach for tackling related issues such as misinformation detection and user behavior analysis.

In this paper, we conceptualize \textit{credibility} as a quantifiable metric representing the reliability and trustworthiness of a news source. Credibility plays a pivotal role in shaping public perception of the media, particularly as social media increasingly shifts the flow of influence from traditional news outlets to content creators and influencers \cite{Amini2024, bayiz2024susceptibility, sitaula2019credibilitybasedfakenewsdetection}. We define misinformation by the contents intentionally or inadvertently altered or fabricated with the specific purpose of influencing or manipulating public perception. Although a high credibility score does not unequivocally confirm the reliability on all information disseminated by a source, it statistically correlates with a diminished probability of disseminating misinformation. 

\begin{figure*}[t]
    \centering
    \includegraphics[width=.8\textwidth]{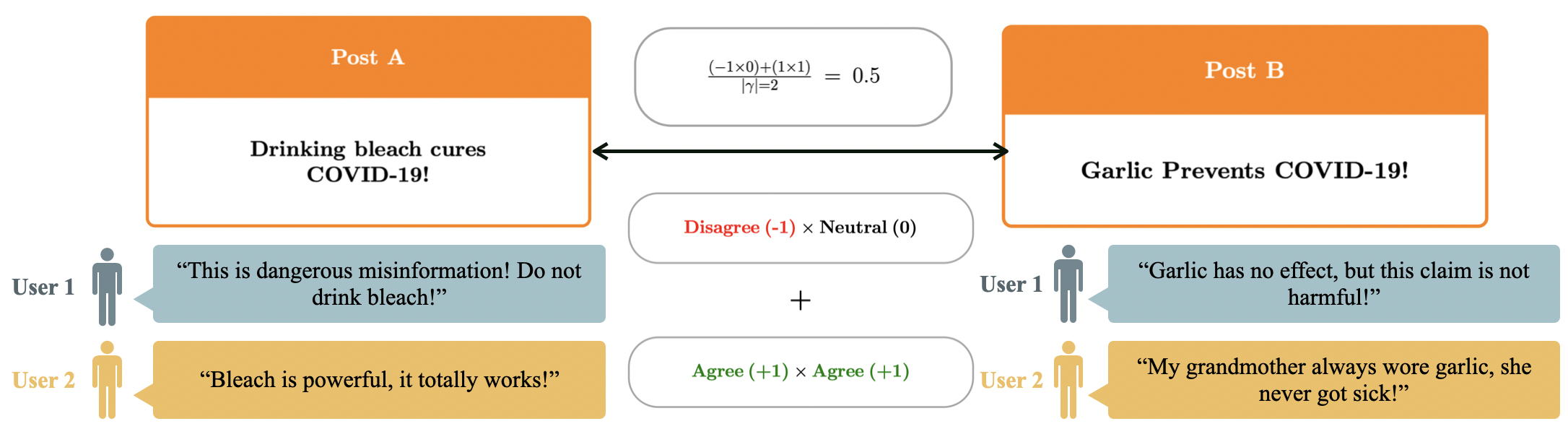}
    \caption{Example of how our post-to-post graph is built. We link posts if they share at least one commenter. The edge weight is determined by taking the product of the stances of each user to the pair of posts and averaging it over all such shared users. This weighting scheme contains non-trivial information that our algorithm \textit{CrediRAG} leverages for effective detection of misinformation across Reddit.}
    \label{fig:post_to_post_graph}
\end{figure*}

In an effort to combat the issues faced by text-only classifiers that ignore the underlying rich network information or graph-based classifiers that are not able to incorporate dynamic real-time data for predictions, we propose \textit{CrediRAG}, a fake news detection method that provides accurate and explainable detection for misinformative content in Reddit. Our approach diverges from traditional fake news detection methodologies by combining two significant architectures. \textbf{First}, we have an explainable search over a continual knowledge base of recent news. \textbf{Second}, this search is then combined with network-based refinement of detected labels leveraging user interactions to construct a post-to-post social graph\footnote{An example of this can be seen in Figure \ref{fig:post_to_post_graph}.}. 

Using real-world Reddit data described in Table \ref{tab:subreddit_distributions_key_single}, we create a weighted post-to-post graph where each edge represents the average stance of shared commenters between pairs of posts. 

We present an innovative approach for retrieval based on source credibility, leveraging dense retrieval techniques that use semantic search of submission embeddings against a comprehensive knowledge base. This enables us to retrieve the top \textit{k} news articles\footnote{In our experiments, we set $k=20$ articles. Details of the implementation are provided in the appendix.} from the \textbf{AskNews}\footnote{The AskNews API information can be found here: \url{https://asknews.app/en}}corpus—a repository of over 50,000 sources, updated every four hours, spanning 14 languages \cite{Tornquist2024}.

We estimate the misinformation label of posts by comparing their sources' average credibility, as determined from the retrieved articles, using the Ad Fontes Media dataset\footnote{The dataset is detailed at \url{https://adfontesmedia.com}. Each source has a credibility score between 0 and 64, which we normalize to a scale of 0 to 1.}. These estimates are then refined through a graph attention network (GAT) pre-trained on community structures. This approach updates the misinformation classification by considering both the content and its position within the subreddit community. Our method surpasses existing graph-based approaches that focus on node-level features \cite{Hurtado2019, Amini2024}, or traditional retrieval-augmented generation (RAG) models \cite{Li2024, Liao2023}, which often overlook the complex social graph of Reddit communities.

Building on the networked structure of these interactions, we seek to address two key research questions:
\begin{description}
\item\textbf{RQ1:} Can retrieval techniques, when integrated with source credibility evaluations, effectively identify misinformative posts on social media? If so, how do they compare in performance to current state-of-the-art approaches?
\item\textbf{RQ2:}
 Can incorporating a Graph Attention Network (GAT) as a post-processing layer that utilizes the interaction data enhance the effectiveness of retrieval-based systems in detecting misinformation? 
\end{description}

These questions are key as they examine how integrating source credibility retrieval and GATs enhances misinformation detection, demonstrating the promise of language models to boost accuracy while addressing the challenge of reduced explainability in complex models. Our main contribution is CrediRAG, a system that uses adversarially trained GATs to refine RAG credibility labels, significantly improving detection accuracy by capturing complex relationships between posts and leveraging generative AI.

The rest of the paper is structured as follows: Section 2 reviews related work and how the proposed framework fills the gaps in fake news detection. Section 3 covers data curation and evaluation datasets. Section 4 explains CrediRAG’s pipeline. We present the results and their implications in Section 5. Sections 6 and 7 discuss broader impacts, limitations, and future research. Experimental details are given in the Appendix.

\section{Related Work}
\subsection{Large Language Models (LLMs) for Fake News Detection}
Large Language Models (LLMs) have shown significant promise in detecting fake news due to their ability to process large amounts of textual data \cite{kim2024llmsproducefaithfulexplanations}. However, they face several limitations, including reliance on static external knowledge sources, vulnerabilities to adversarial attacks, and challenges with cross-domain generalization \cite{li_2023, wu2023cheapfakedetection, guan2024languagemodelshallucinateexcel, liu2024raemollmretrievalaugmentedllms, hang2024trumorgpt}. For example, knowledge-enhanced models \cite{whitehouse_2022} and modular architectures like \textit{Self-Checker} \cite{li_2023} improve detection by integrating external databases but are limited by the quality and availability of external knowledge, reducing their real-time efficiency.

Other studies have explored prompt engineering techniques to optimize LLMs, as seen in the detection of "cheap-fakes" \cite{wu2023cheapfakedetection}, though such approaches struggle with more complex forms of disinformation like deepfakes. Additionally, the dual-use nature of LLMs raises ethical concerns, as they can be exploited to generate misinformation \cite{hu2024badactor}. Fine-tuning and weak supervision methods \cite{pavlyshenko2023analysisdisinformationfakenews, leite2023detectingmisinformationllmpredictedcredibility} enhance LLMs' domain-specific performance, but come with significant computational costs and inconsistencies in handling noisy, real-world data.

A key issue with LLMs is their propensity to "hallucinate" facts, generating incorrect information \cite{guan2024languagemodelshallucinateexcel}. Moreover, these models are susceptible to adversarial attacks such as clean-label poisoning \cite{liang2023cleanlabel}, further undermining their reliability. While hybrid models that combine human intelligence and LLMs show promise \cite{yang2023rumor}, they suffer from potential biases and decreased accuracy in dynamic social media environments \cite{liu2024largelanguagemodelsdetect}.

Cross-domain generalization remains a significant challenge. Retrieval-augmented frameworks like RAEmoLLM \cite{liu2024raemollmretrievalaugmentedllms} have integrated emotional cues for improved misinformation detection, but this introduces risks of misclassification when emotional intensity does not align with truthfulness. Query-based fact-checking models like "TrumorGPT" \cite{hang2024trumorgpt} rely heavily on query quality, making them less reliable when faced with poorly formulated queries.

\subsection{Graph Neural Networks (GNNs) for Fake News Detection}
Graph Neural Networks (GNNs) have gained prominence in detecting fake news by modeling the complex relationships among users, content, and interactions. Despite their strong performance, GNNs face significant limitations related to explainability, scalability, and sensitivity to dynamic social media environments. While continual learning \cite{han2020graph} and geometric deep learning \cite{monti2019fake} can mitigate model degradation over time and improve accuracy, the increased complexity limits models' interpretability. Bi-directional Graph Convolutional Networks (GCNs) \cite{bian2020rumor} further enhance rumor detection by capturing the spread and propagation of information, though their decision-making processes remain opaque, challenging public trust.

Key GNN architectures, such as Graph Convolutional Networks (GCNs) \cite{kipf2016semi} and Graph Attention Networks (GATs) \cite{velickovic2017graph}, allow models to prioritize influential nodes, but attention mechanisms and large-scale graph structures introduce scalability issues, with models struggling to handle vast and intricate social networks. Inductive representation learning \cite{hamilton2017inductive} addresses some scalability concerns but also sacrifices interpretability, as abstract representations remain difficult to deconstruct, making models' decisions unclear.

GNN-based models also face challenges in dynamic, adversarial misinformation environments. While approaches like edge confidence evaluation met for detecting bots \cite{qiao2024dispelling} enhance detection accuracy by assessing the reliability of social connections, the complexity of these evaluations reduces explainability, which, combined with the dynamic nature of misinformation environments, makes trusting model outputs difficult without human oversight. Heterogeneous Subgraph Transformers \cite{zhang2024heterogeneoussubgraphtransformerfake} improve detection by accounting for diverse interactions, but increased heterogeneity complicates the model’s structure, reducing transparency and limiting real-time scalability.

Text-based GNN models like TCGNN \cite{li2024tcgnn} improve detection by clustering similar content before applying graph-based techniques. However, this approach introduces abstraction layers that reduce transparency, and errors in clustering can decrease overall reliability in misinformation detection.

\subsection{Contributions}
This paper distinguishes itself from the existing body of work with the following three contributions:
\begin{itemize}
    \item \textbf{Novel Integration of Graph Community Structures and Credibility-Driven Retrieval}: We are the first to leverage Reddit's intricate graph community structure and dynamic credibility-driven retrieval for more accurate fake news detection. This approach uniquely exploits the relational dynamics within social media platforms to enhance detection performance.
    
    \item \textbf{CrediRAG Model with Post-to-Post Network Architecture}: We propose the \textit{CrediRAG} model which improves accuracy and enhances explainability. By assigning initial credibility estimates based on similar news articles and improving them by incorporating a novel post-to-post network, our model captures the graph-level information often overlooked by text-based classifiers while providing the explainability behind the decision-making often missed in traditional graph-based methods. 
    
    \item \textbf{Curated Fake News Dataset}: We introduce a new labeled dataset of over 60,000 Reddit posts from 2016-2018, curated by matching verified fake news titles from the ISOT Dataset \cite{ahmed2017detection, ahmed2018detecting} to Reddit submissions. This dataset offers a valuable resource for advancing graph-based misinformation research and enhancing the detection of fake news across social platforms.
\end{itemize}

\section{Data}
Given the lack of substantially labelled Reddit data, for our experiments, we created our own labelled dataset\footnote{We will make this dataset public for encouraging further graph-based research in fake news detection on social media.} as described below. We also validated our results through experiments on the \textit{r/Fakeddit} dataset \cite{Nakamura2020}.

\subsection{Dataset Construction}
We develop a novel approach by leveraging the ISOT Fake News Dataset\footnote{Note that the dataset can be found \url{https://www.kaggle.com/datasets/emineyetm/fake-news-detection-datasets}. It is from Dr. Hadeer Ahmed from the University of Victoria \cite{ahmed2017detection, ahmed2018detecting}.} and aligning it with unlabeled subreddit data extracted from Pushshift.io\footnote{\url{https://pushshift.io}} \cite{ahmed2017detection, ahmed2018detecting, baumgartner2020pushshiftredditdataset}. Our goal is to enrich the dataset by matching Reddit submissions to known fake and true news articles from the ISOT dataset. This was achieved through a two-step process. First, we collect Reddit submissions and comments and then subsequently match them with fake and true news articles based on temporal proximity and textual similarity.

\subsubsection{Subreddit Selection}
To ensure that our dataset captures relevant discussions prone to misinformation, we carefully select subreddits that frequently discuss controversial topics, especially politics and social issues. Some of the top subreddits included in our data are shown in Table \ref{tab:subreddit_distributions_key_single}.

These communities are selected because they frequently involve discussions surrounding political ideologies, health misinformation (e.g., COVID-19, anti-vaccination), and global events (e.g., the war in Ukraine). They are also known to be common spaces for both legitimate and false information. Note all the data is in English only.

\subsubsection{Data Preprocessing}
Initially, we filter the data by comment number to retain only the most significant posts for analysis. For each submission, we then compare the publication date of the Reddit post with the dates of ISOT articles. Submissions and comments that fell within a predefined window of time---two days ---around the publication dates of fake or true news articles were retained for further matching. This process allows us to filter out irrelevant data and focus only on posts that were temporally aligned with news events.

\subsubsection{News Matching}
To match Reddit submissions and ISOT news articles, we employ the \textit{SentenceTransformer} model, specifically the \texttt{all-MiniLM-L6-v2} architecture \cite{reimers2019sentencebert}. This model generates dense vector representations of the text input. Given a title from a Reddit post, represented as text input $T_r$, the model produces an embedding $\mathbf{A}$ for $T_r$. Similarly, it generates a high dimensional latent space embedding $\mathbf{B}$ for a news article title $T_n$. The model optimizes for embeddings such that semantically similar text inputs (e.g., titles discussing the same event) have closer representations in this vector space. We compute the cosine similarity between article title embeddings $\mathbf{A}$ and $\mathbf{B}$ as follows:
\begin{align}
S(\mathbf{A}, \mathbf{B}) = \frac{\mathbf{A} \cdot \mathbf{B}}{\|\mathbf{A}\| \|\mathbf{B}\|},
\end{align}
where $\mathbf{A} \cdot \mathbf{B}$ is the dot product of the two vectors, and $\|\mathbf{A}\|$ and $\|\mathbf{B}\|$ represent the $2$-norm of the vectors ${\bf A},~{\bf B}$ respectively. Cosine similarity yields a score between $-1$ and $1$, where $1$ indicates perfect similarity. In our implementation, we set a similarity threshold of $0.7$, meaning that only submissions whose titles have a similarity score greater than or equal to $0.7$ with a news article title are considered matched.
The final dataset consists of posts matched with fake or true news articles, labeled based on the corresponding news article. The proposed method is conservative, matching posts with news only when both textual and temporal similarities are met. Posts that do not match any news article in the dataset are discarded.

\subsection{Dataset Statistics}
We illustrate the specifics of our top subreddits in Table \ref{tab:subreddit_distributions_key_single}. Note that we train our corrective Graph Attention Network detailed in the methodology section (Section 4) on \textit{r/Conservative} and \textit{r/Democrats}, and test it on \textit{r/Libertarian} and \textit{r/Impeach Trump} from our curated dataset. On the r/Fakeddit data, we train on their entire text-only training data containing around $100,000$ posts and test on their entire text-only testing data containing around $8,000$ posts.
\begin{table}
\centering
\caption{Distribution of Fake and Real News Posts in Key Political Subreddits in our curated dataset (2016-2018)}
\small
\begin{tabular}{@{}lcc@{}}
\toprule
\textbf{Subreddit} & \textbf{Fake News Entries} & \textbf{Real News Entries} \\
\midrule
r/Conservative & $3952$ & $2250$ \\
r/EnoughTrumpSpam & $5521$ & $3462$ \\
r/ImpeachTrump & $1077$ & $856$ \\
r/Democrats & $1439$ & $1020$ \\
r/News & $20957$ & $21859$ \\
r/Libertarian & $857$ & $689$ \\
r/Progressive & $620$ & $481$ \\
r/Republican & $852$ & $685$ \\
r/Conspiracy & $4715$ & $2415$ \\
r/Politics & $48108$ & $56221$ \\
r/DonaldTrumpWhiteHouse & $3098$ & $4313$ \\
r/SandersForPresident & $2054$ & $1417$ \\
\bottomrule
\end{tabular}
\label{tab:subreddit_distributions_key_single}
\end{table}

\section{Methodology}
We present a robust pipeline for classifying misinformation on Reddit through the combination of graph networks with information retrieval.  Figure \ref{fig:methodology} provides a high-level overview of the process.


\subsection{Retrieval-Augmented Generation (RAG)}
Accurate detection of fake news is a challenging task, particularly in dynamic environments like social media, where information is rapidly produced and shared. Traditional approaches to fake news detection often rely on static classifiers or models trained on manually curated datasets. However, these methods can struggle with crude labeling practices and limited context, leading to reduced accuracy in real-world scenarios. To address these limitations, we use RAG \cite{lewis2021retrievalaugmentedgenerationknowledgeintensivenlp}, which enhances detection by incorporating external knowledge dynamically during inference. RAG offers a significant advantage by retrieving relevant information in real-time. 

We use the \textit{AskNews} corpus to dynamically retrieve relevant similar news articles to the submission body of a post \cite{Tornquist2024}. The AskNews API provides an efficient way to integrate real-time news into LLM applications. Using its RAG architecture, AskNews processes over 300,000 articles daily, embedding them into a vector database that can be queried with natural language.

During inference, our model retrieves the top-$k$ news articles that semantically align with the test post, ensuring that the model’s predictions are informed by up-to-date and contextually relevant content. Our RAG approach mitigates the weaknesses of crude labeling and static data reliance, offering a more robust and context-aware solution for fake news detection.

Formally, let $\mathcal{D}$ represent the corpus of news articles and $\mathbf{q} \in \mathbb{R}^d$ represent the query vector encoding the post's content, where $d$ is the dimensionality of the vector space. We select the top-$k$ most similar documents by maximizing the similarity between the query vector $\mathbf{q}$ and document vectors $\mathbf{d} \in \mathcal{D}$. The similarity \textbf{sim} is computed using the dot product: 
\begin{align}
\text{sim}(\mathbf{q}, \mathbf{d}) = \mathbf{q}^\top \mathbf{d}.
\end{align}
For efficiency, we use an approximate nearest neighbor (ANN) search, which allows us to avoid performing an exhaustive search over the entire corpus $\mathcal{D}$. The goal is to retrieve the top-$k$ documents by solving the following optimization problem:
\begin{align}
E = \operatorname*{argmax}_{d_i \in \mathcal{D}, |E| = k} \left( \mathbf{q}^\top \mathbf{d} \right),
\end{align}

where $E$ is the set of top-$k$ documents that maximize the inner product with the query vector $\mathbf{q}$.

Once the top-$k$ articles are retrieved, we evaluate their credibility using predefined credibility scores from the Ad Fontes Media\footnote{This approach of News Source Credibility using Ad Fontes is similar to other papers \cite{Amini2024,bayiz2024susceptibility}}, based on the news source of each document. Let $C(\Lambda(d_i))$ represent the credibility score of the source $\Lambda(d_i)$ of document $d_i$. We aggregate the credibility of the retrieved documents by computing the predicted label $\hat{y}$ as the mean of the credibility scores of the retrieved articles; that is:

\begin{align}
\hat{y} = \frac{\sum_{i=1}^k C(\Lambda(d_i))}{k}.
\end{align}

\subsection{Label Refinement}
Despite their high accuracy in fake-news detection tasks, retrieval-based methods suffer from several problems that limit their usefulness in a social network setting \cite{liu2024raemollmretrievalaugmentedllms}. Firstly, these methods only take the content of the message into account, therefore they omit the underlying user interactions in their label assignment. Such interactions can be highly informative in determining fake news as they contain information about the community structure and the echo chambers within the social network. Secondly, some of the posts may contain text that is either too short or absent entirely. For instance, if the post itself is an image, the title itself may not contain sufficient information to determine the credibility of the entire content.

To alleviate these issues and improve prediction accuracy, we employ a refinement step of a graph attention network over the post-to-post networks. Intuitively, this step corrects the errors in the initial credibility assignments made in the retrieval-based generation step using the post-to-post network as side information.

\subsubsection{Post-to-Post Network Construction}
Given a set of posts labeled as \texttt{real\_news} or \texttt{fake\_news}, we construct a graph where nodes represent individual posts, and edges capture the relationship between posts that share commenters. Formally, let $\mathcal{P} = \{P_1, P_2, \dots, P_n\}$ represent the set of posts, and $\mathcal{C} = \{C_1, C_2, \dots, C_m\}$ represent the set of commenters. An undirected edge $e_{ij}$ exists between two posts $P_i$ and $P_j$ if they share at least one commenter, i.e., if there exists $C_k$ such that $C_k \in P_i$ and $C_k \in P_j$. This procedure is described in Algorithm~\ref{alg:post2post}.

The edge weights between two posts $P_i$ and $P_j$ are determined using Algorithm~\ref{alg:weightedge}. For each shared commenter $C_k$ between posts $P_i$ and $P_j$, we calculate the stance $\sigma(C_k, P_i)$ and $\sigma(C_k, P_j)$, which represents whether the commenter agrees or disagrees with the content of each post. Let $\sigma(C_k, P_i), \sigma(C_k, P_j) \in \{-1, 0, 1\}$ represent \texttt{disagree}, \texttt{neutral}, and \texttt{agree}, respectively. We compute the edge weight as:

\begin{align}
\text{weight}(P_i, P_j) = \frac{1}{|\mathcal{\gamma}|} \sum_{C_k \in \mathcal{\gamma}} \sigma(C_k, P_i) \times \sigma(C_k, P_j),
\end{align}

where $\mathcal{\gamma}$ is the set of shared commenters between $P_i$ and $P_j$. The detailed edge weight computation is outlined in Algorithm~\ref{alg:weightedge}, while the overall network construction process is described in Algorithm~\ref{alg:post2post}. Note that an example graph can be found in Figure \ref{fig:post_to_post_graph}.

\begin{algorithm}[H]
\caption{Post2Post Network}
\label{alg:post2post}
\begin{algorithmic}[1]
    \FOR{all pairs of posts $A$ and $B$ labeled \textit{fake news} or \textit{real news}}
        \IF{$A$ and $B$ share a commenter}
            \STATE Create an edge between $A$ and $B$
            \STATE Call \texttt{WeightEdge($A$, $B$)} to assign a weight to the edge
        \ENDIF
    \ENDFOR
\end{algorithmic}
\end{algorithm}

\begin{algorithm}[H]
\caption{WeightEdge($A$, $B$)}
\label{alg:weightedge}
\begin{algorithmic}[1]
    \STATE Initialize weight to 0
    \FOR{each shared commenter $C_1$ between posts $A$ and $B$}
        \STATE Compute $\sigma(C_1, A)$ and $\sigma(C_1, B)$
        \STATE $\sigma(C_1, A), \sigma(C_1, B) \in \{-1, 0, 1\}$ representing disagree, neutral, and agree, respectively
        \STATE weight $\gets$ weight + $\sigma(C_1, A) \times \sigma(C_1, B)$
    \ENDFOR
    \STATE weight $\gets$ weight / $|\mathcal{\gamma}|$
\end{algorithmic}
\end{algorithm}

\subsubsection{Graph Attention Networks}

Graph Neural Networks (GNNs) have become a powerful tool for modeling relational data, such as social networks, citation networks, and knowledge graphs, as demonstrated in various studies \cite{han2020graph, zhang2024heterogeneoussubgraphtransformerfake, velickovic2017graph, kipf2016semi, bian2020rumor, monti2019fake}. At their core, GNNs propagate information through a graph by leveraging the graph's connectivity structure to improve node-level predictions. One commonly used GNN model is the Graph Convolutional Network (GCN), which updates a node’s representation by aggregating features from its neighbors. For a graph \( G = (V, E) \), where \( V \) is the set of nodes and \( E \) is the set of edges, the update rule for each node \( v_i \in V \) in a GCN can be written as:

\begin{align}
\mathbf{h}_i^{(l+1)} = \phi\left( \sum_{j \in \mathcal{N}(i)} \frac{1}{\sqrt{d_i d_j}} \mathbf{W}^{(l)} \mathbf{h}_j^{(l)} \right),
\end{align}

where \( \mathbf{h}_i^{(l)} \) is the feature vector of node \( i \) at layer \( l \), \( \mathbf{W}^{(l)} \) is a learnable weight matrix, \( \mathcal{N}(i) \) is the set of neighbors of node \( i \), \( d_i \) is the degree of node \( i \), and \( \phi \) is an activation function, such as ReLU. However, this aggregation process in GCNs treats all neighbors equally, weighted only by their degree. This is a limitation when some neighbors are more relevant or influential than others, particularly in heterogeneous graphs like ours, where certain posts are more influential than others.

Given that our dataset contains posts of varying importance based on their connectivity and credibility, we require a model that can dynamically adjust the contribution of each neighbor. To address this, we adopt Graph Attention Networks (GATs), which use attention mechanisms to weigh the importance of neighboring nodes.

Once the post-to-post graph is constructed, we train a GAT to classify each node (post) as either \texttt{real\_news} or \texttt{fake\_news}. Let \( G = (V, E) \) represent our graph, where \( V \) is the set of posts and \( E \) is the set of weighted edges. For each node \( v \in V \), the GAT computes the attention coefficient \( \alpha_{ij} \) for every neighboring node \( v_j \in \mathcal{N}(v) \) as follows:

\begin{align}
\alpha_{ij} = \frac{\exp \left( \text{LeakyReLU}\left( \vec{a}^\top \left[\mathbf{W} \mathbf{h}_i \parallel \mathbf{W} \mathbf{h}_j\right]\right)\right)}{\sum_{k \in \mathcal{N}(i)} \exp \left( \text{LeakyReLU}\left( \vec{a}^\top \left[\mathbf{W} \mathbf{h}_i \parallel \mathbf{W} \mathbf{h}_k\right]\right)\right)},
\end{align}

where \( \mathbf{h}_i \) is the feature representation of node \( i \), \( \mathbf{W} \) is a learnable weight matrix, \( \vec{a} \) is the attention vector, and \( \parallel \) represents vector concatenation. The final node representation is a weighted sum of its neighbors:

\begin{align}
\mathbf{h}_i' = \phi\left( \sum_{j \in \mathcal{N}(i)} \alpha_{ij} \mathbf{W} \mathbf{h}_j \right),
\end{align}

where \( \phi \) is the activation function. This attention mechanism allows the GAT to dynamically focus on more relevant or credible neighbors, giving the model a greater ability to distinguish between real and fake news.

During training, we need a large set of posts with both ground truth credibility labels and initial credibility assignments from the retrieval-based method. However, querying a retrieval model for each post is computationally expensive. To mitigate this, we generate fictitious initial credibility assignments by randomly corrupting a subset of the labels and training the GAT to reconstruct the original ground truth. This process aims to improve the model’s robustness by learning to correct errors in the initial labeling.

The ratio \( r \) of corrupted labels must be tuned to reflect the false classification rate of the retrieval-based classifier. In our experiments, we found that setting \( r = 15\% \) yielded optimal results, based on preliminary evaluations of the retrieval-based classifier.

\section{Results and Discussion}
\begin{table*}
    \centering
    \caption{Model Performance on ISOT Reddit and r/Fakeddit Datasets. We train all models on 8,661 nodes and test on 3,479 for ISOT Reddit, and train on 100,000 nodes and test on 8,000 nodes for r/Fakeddit. Shown in bold is our main CrediRAG model.}
    \begin{adjustbox}{max width=\textwidth}
    \begin{tabular}{@{}lcccccc@{}}
        \toprule
        \textbf{Method} & \textbf{Model} & \multicolumn{2}{c}{\textbf{ISOT Reddit}} & \multicolumn{2}{c}{\textbf{r/Fakeddit}} \\
        \cmidrule(lr){3-4} \cmidrule(lr){5-6}
        & & \textit{Accuracy} & \textit{F1-Score} & \textit{Accuracy} & \textit{F1-Score} \\
        \midrule
        \multirow{2}{*}{\textit{Random}} & Random & 0.355 & 0.402 & 0.300 & 0.350 \\
        & Majority & 0.551 & 0.433 & 0.500 & 0.400 \\
        \midrule
        \multirow{2}{*}{\textit{Neural Network-Based}} & TextCNN \cite{kim2014convolutionalneuralnetworkssentence} & 0.598 & 0.547 & 0.620 & 0.570 \\
        & TextGCN \cite{yao2018graphconvolutionalnetworkstext}& 0.623 & 0.644 & 0.650 & 0.610 \\
        & GAT-only \cite{velickovic2017graph} & 0.650 & 0.670 & 0.660 & 0.675 \\
        \midrule
        \multirow{3}{*}{\textit{Transformer-Based}} & BERT \cite{devlin2019bertpretrainingdeepbidirectional} & 0.624 & 0.667 & 0.700 & 0.690 \\
        & GPT3.5-turbo \cite{brown2020languagemodelsfewshotlearners} & 0.678 & 0.677 & 0.715 & 0.710 \\
         & CrediBERT \cite{Amini2024} & 0.677 & 0.699 & 0.720 & 0.715 \\
        \midrule
          \multirow{2}{*}{\textit{RAG-Based}}& Web-Retrieval LLM Agent \cite{tian2024web} & 0.678 & 0.711 & 0.730 & 0.720 \\
         & CrediRAG without GAT (Ours) & 0.732 & 0.722 & 0.750 & 0.740\\
        \midrule
         \multirow{2}{*}{\textit{GNN + RAG}} & CrediRAG with un-weighted GAT (Ours) & 0.764 & 0.766 & 0.850 & 0.890 \\
         & \textbf{CrediRAG with Weighted GAT (Ours)} & \textbf{0.8667} & \textbf{0.811} & \textbf{0.9437} & \textbf{0.920} \\
        \bottomrule
    \end{tabular}
    \end{adjustbox}
    \label{experiment_results}
\end{table*}
We present the performance of various models across both the \textit{ISOT Reddit} and \textit{r/Fakeddit} datasets in Table \ref{experiment_results}. Here, \textbf{accuracy} refers to the proportion of correctly classified posts (both real and fake) out of the total number of posts, while \textbf{F1-score} gives a better indication of the model's performance in identifying fake news posts especially considering class imbalance. Our proposed model, \textit{CrediRAG with Weighted GAT}, consistently outperforms all other baselines in terms of accuracy and F1-score across both datasets. These ablation results highlight the effectiveness of our design choices in CrediRAG, particularly in leveraging graph-based attention mechanisms and retrieval-augmented generation for detecting fake news on social media.

\subsection{Performance on ISOT Reddit}
As shown in Table \ref{experiment_results}, on the \textit{ISOT Reddit} dataset, CrediRAG with Weighted GAT achieves the highest accuracy of \textit{0.8667} and an F1-score of \textit{0.811}, outperforming all other models. The next best model, \textit{CrediRAG with un-weighted GAT}, achieves an accuracy of \textit{0.764} and an F1-score of \textit{0.766}. This clear improvement of approximately \textit{10\%} in accuracy and \textit{4.5\%} in F1-score demonstrates the critical impact of introducing weighted edges in the graph attention mechanism. Through these ablation tests, we reveal that the improvement in performance is due to our various architectural choices in designing \textit{CrediRAG}.

We also evaluated a variant of our model, \textit{CrediRAG without GAT}, which omits the graph attention component. This variant achieves an accuracy of \textit{0.732} and an F1-score of \textit{0.722}, indicating a substantial decline in performance compared to models utilizing graph-based attention. These results illustrate that while RAG contributes to performance gains, the inclusion of GAT layers is essential for capturing the complex relationships between posts and comments within Reddit discussions. The GAT layers enable the model to learn more effectively from the network structure of the data, allowing it to better classify posts as misinformative or not. In contrast, approaches like Google-search based retrieval models \cite{tian2024web} are limited in their ability to generalize to complex and dynamic discussions on platforms like Reddit. Google search accesses content from a broader web index and retrieves documents with little consideration of the conversation's context or the interactions between users. This makes it less effective for detecting misinformation within highly interactive, user-generated environments like Reddit, where understanding the relationships between posts, replies, and user behaviors is crucial. By integrating the GAT into our RAG framework, \textit{CrediRAG} offers superior performance in both classification and generalization, successfully leveraging both content and relational structures within social media platforms.

\subsection{Performance on r/Fakeddit}
On the \textit{r/Fakeddit} dataset, the performance of CrediRAG with Weighted GAT is even more pronounced, with an accuracy of \textit{0.9437} and an F1-score of \textit{0.920}, once again demonstrating a significant improvement over other baselines. The next best model, \textit{CrediRAG with un-weighted GAT}, achieves an accuracy of \textit{0.850} and an F1-score of \textit{0.890}, confirming that the weighted GAT mechanism captures more nuanced interactions between nodes in the graph and reinforcing the impact of our post-to-post network weighting structure.

It is important to note that the optimistic performance on \textit{r/Fakeddit} may partially stem from the dataset's labeling scheme. In r/Fakeddit, all posts from certain subreddits are pre-labeled as either fake or real. This subreddit-wide labeling may artificially inflate the accuracy and F1-score, as models can exploit these global patterns rather than focusing solely on the content of individual posts. Therefore, while the performance improvements on r/Fakeddit are significant, the labeling approach should be considered when interpreting these results.

\subsection{Significance of the Results}
The results in Table \ref{experiment_results} demonstrate that integrating GNN-based attention and retrieval-augmented generation in \textit{CrediRAG} significantly outperforms models like \textit{BERT}, \textit{GPT3.5-turbo}, \textit{TextCNN}, and \textit{TextGCN}. While \textit{BERT} achieves an accuracy of 0.624 and an F1-score of 0.667 on ISOT Reddit, \textit{CrediRAG with Weighted GAT} improves accuracy by 24\% and F1-score by 14.4\%. This highlights the importance of graph-based attention in modeling user interactions and post-comment relationships.

The enhanced performance of \textit{CrediRAG with Weighted GAT} underscores the value of edge weights in GAT layers. Assigning varying importance to user interactions improves the model's effectiveness. In contrast, \textit{CrediRAG without GAT} demonstrates the critical role of graph-based attention for detecting misinformation.

\subsection{Case Study: Generalizability of \textit{CrediRAG}}

\begin{figure*}[ht!]
    \centering
    \includegraphics[width=0.24\textwidth]{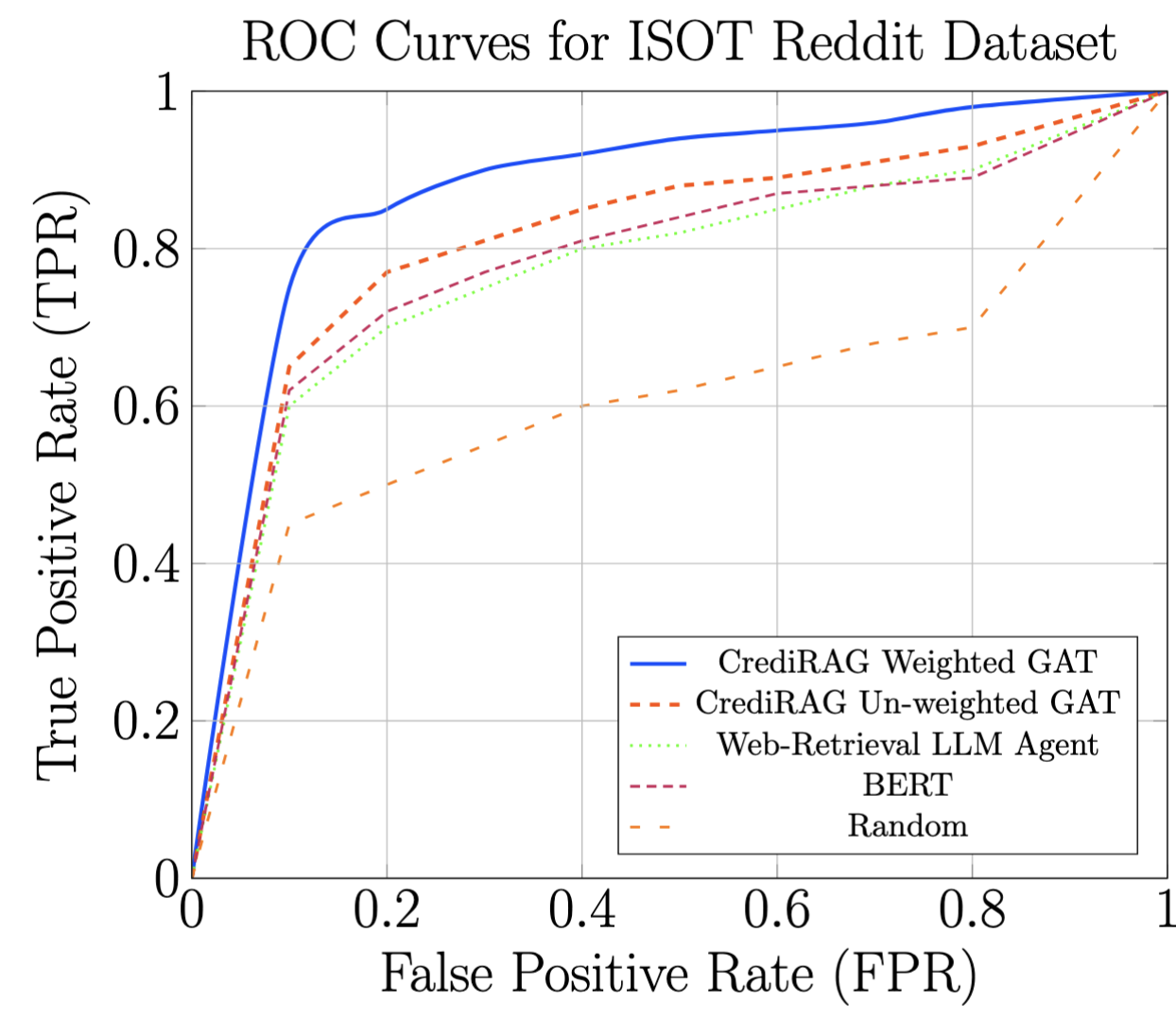}
    \includegraphics[width=0.24\textwidth]{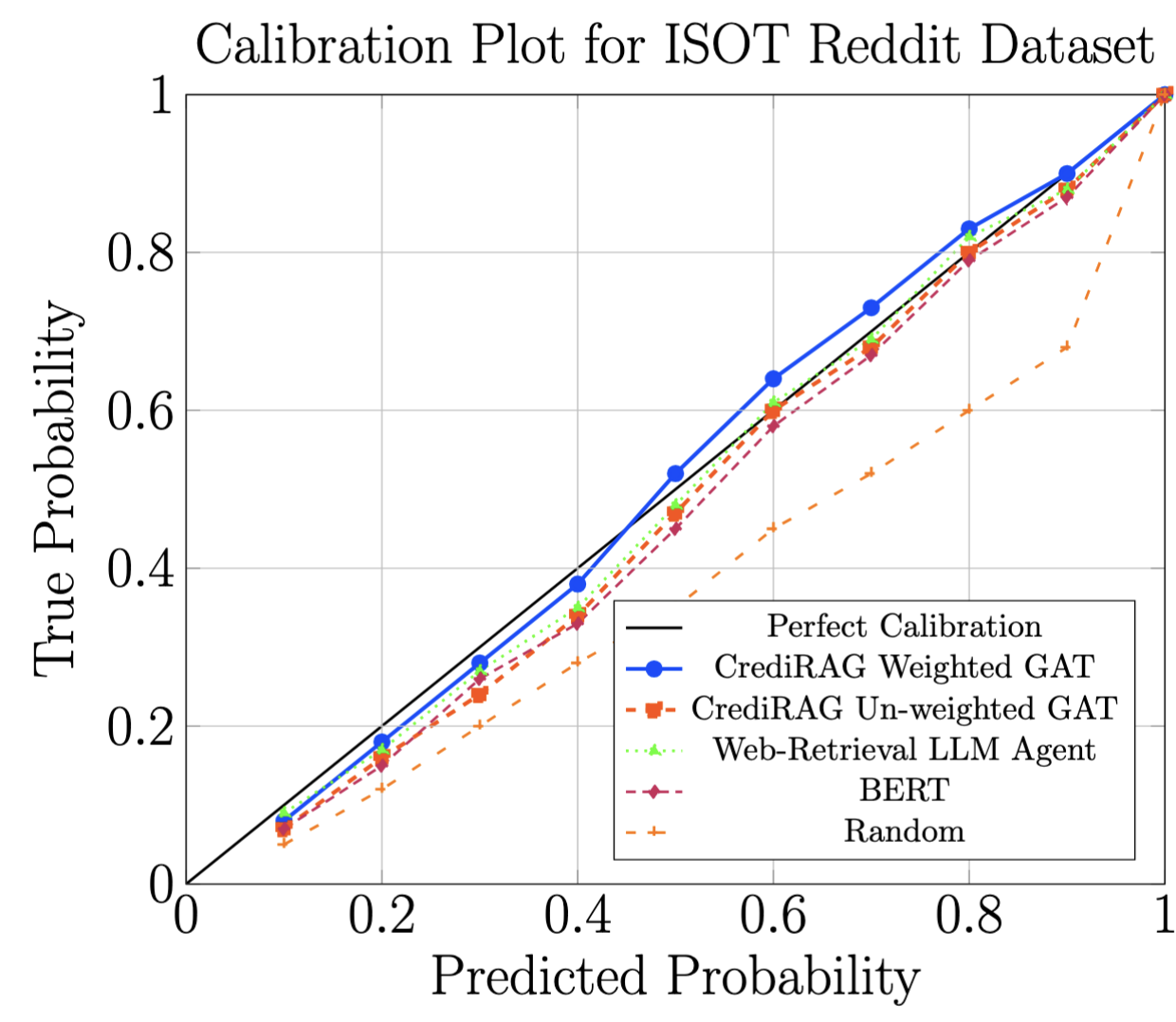}
    \includegraphics[width=0.24\textwidth]{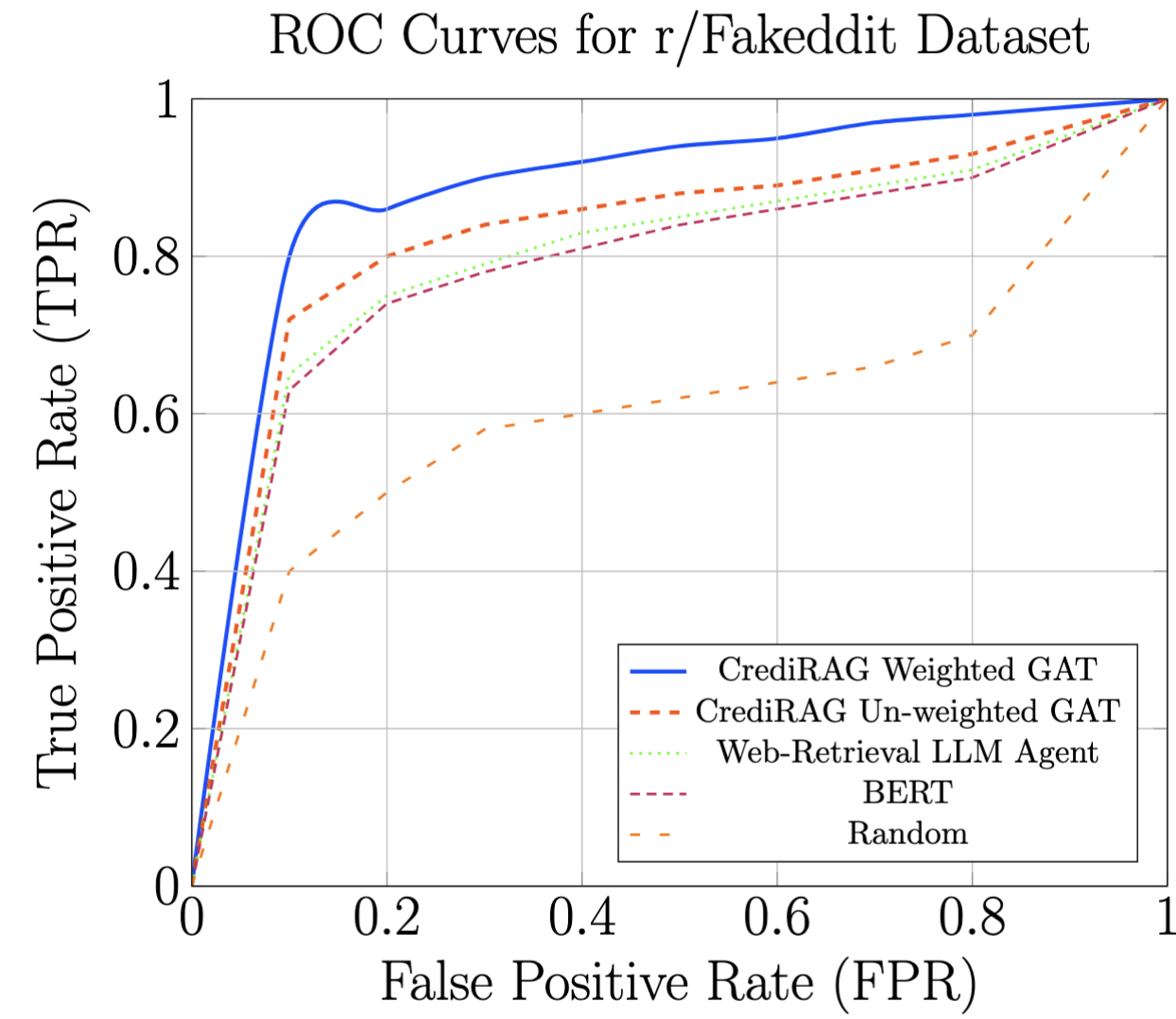}
    \includegraphics[width=0.24\textwidth]{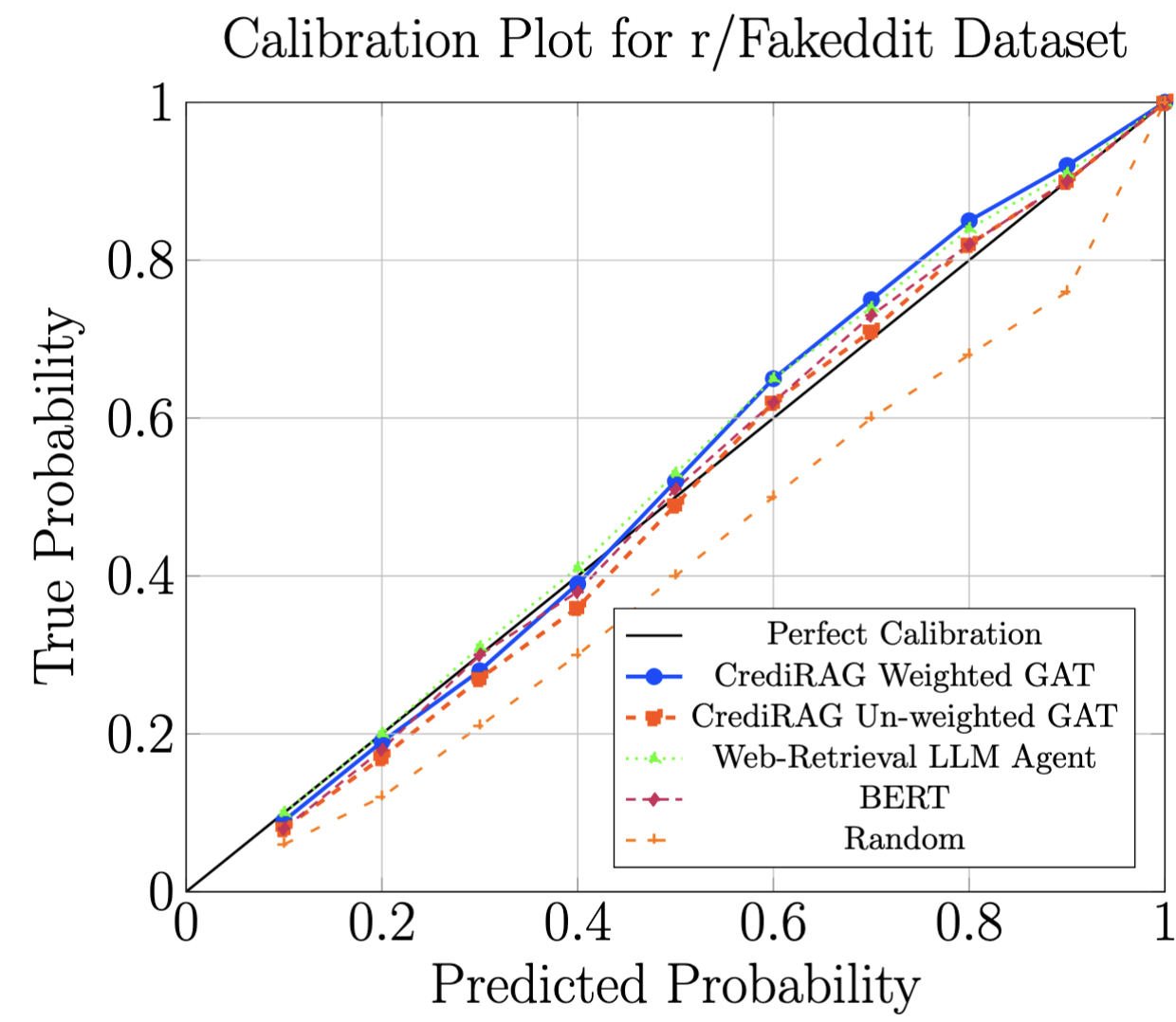}
    \caption{ROC and Calibration Curves for ISOT Reddit and r/Fakeddit Datasets on r/SandersForPresident, r/EnoughTrumpSpam, and r/DonaldTrumpWhiteHouse subreddits. As shown, CrediRAG (blue curve) gets the best results on the ROC curve, while also being well-calibrated.}
    \label{fig:all_figures}
\end{figure*}

In addition to ablation tests, we perform a case study using three subreddits to evaluate the generalizability of \textit{CrediRAG}. Figure \ref{fig:all_figures} shows ROC and calibration curves for the models trained on both \textit{ISOT Reddit} and \textit{r/Fakeddit} datasets, evaluated on \textit{r/SandersForPresident}, \textit{r/EnoughTrumpSpam}, 
and \textit{r/DonaldTrumpWhiteHouse}.

\subsubsection{ROC Curves: Discriminative Power}

The ROC (Receiver Operating Characteristic) curves assess the ability of the model to discriminate between true and false classifications by plotting the true positive rate (TPR) against the false positive rate (FPR). A higher Area Under the Curve (AUC) indicates better performance. 

As shown in the ROC curves (Figure \ref{fig:all_figures}), \textit{CrediRAG with Weighted GAT} demonstrates superior AUC scores on both the \textit{ISOT Reddit} and \textit{r/Fakeddit} datasets, significantly outperforming the baseline models such as \textit{BERT} and \textit{Web-Retrieval LLM Agent}. This steep rise in TPR with minimal increase in FPR indicates that \textit{CrediRAG} is particularly adept at distinguishing between fake and real news, even across unseen subreddits.

\subsubsection{Calibration Plots: Confidence in Predictions}
The calibration plots evaluate the reliability of the model’s predicted probabilities by comparing them to the actual outcomes. A well-calibrated model should closely follow the diagonal line, where predicted probabilities match observed frequencies.

As seen in the calibration plots, \textit{CrediRAG with Weighted GAT} is well-calibrated on both datasets, with most points aligning with the perfect calibration line. In contrast, models like \textit{BERT} and \textit{Random} deviate substantially from the diagonal, indicating overconfidence or underconfidence in their predictions. The accurate probability estimates provided by \textit{CrediRAG} are crucial in real-world applications such as content moderation, where both accuracy and model confidence are necessary for decision-making. 

\subsubsection{Significance of Results}
The results, illustrated by both ROC and calibration plots, show that \textit{CrediRAG with Weighted GAT} not only excels at distinguishing between fake and real news but also provides well-calibrated confidence estimates, making it a robust choice for fake news detection. The model’s generalizability across subreddits further demonstrates its adaptability to unseen data, a critical factor for real-world deployment in misinformation detection.

While \textit{r/Fakeddit}'s optimistic results can be attributed to the dataset's labeling scheme—where entire subreddits are classified as either fake or real—\textit{CrediRAG}'s superior performance on the more granular \textit{ISOT Reddit} dataset highlights that its performance is not merely the result of exploiting global patterns but is driven by its advanced design. This makes \textit{CrediRAG} well-suited for detecting misinformation in diverse Reddit communities.

\section{Broader Impacts}
The advancements of \textit{CrediRAG} have significant implications for fake news detection on platforms like Reddit. By combining graph-based attention with retrieval-augmented generation, \textit{CrediRAG} models both the content and relational structure of discussions, crucial for understanding how misinformation spreads through user interactions and threads.

Our results highlight the potential of integrating GNNs with retrieval techniques for more effective misinformation detection. Additionally, our findings emphasize the need for robust dataset labeling, as simplistic labels (e.g., from r/Fakeddit) can inflate performance by revealing exploitable global patterns. To support further research, we open-source both our code and labeled dataset for graph-based fake news detection on Reddit.

\subsection{Ethics Statement}
Our work uses publicly available Reddit data in compliance with Reddit's terms of service, ensuring user anonymity by not collecting or storing any personally identifiable information. While CrediRAG enhances misinformation detection through source credibility and generative AI, low-credibility outlets may still influence scores. Though our graph-based refinement reduces this risk, we advise users to use the tool cautiously, recognizing its limitations. It is intended as a supplementary resource, and its results should be considered alongside broader assessments of context and credibility.

\section{Conclusion}
In this work, we introduce \textit{CrediRAG}, a novel model that combines RAG with a GAT and a distinct post-to-post structure for the task of fake news detection on Reddit. Through comprehensive evaluations on both the \textit{ISOT Reddit} and \textit{r/Fakeddit} datasets, we demonstrate that \textit{CrediRAG} consistently outperforms various baseline models, achieving superior accuracy, F1-scores, and robust generalization to unseen subreddits. The ROC and calibration curves further underscore \textit{CrediRAG}'s ability to provide both high discriminative power and reliable confidence estimates. 

Our results validate the importance of incorporating graph-based attention mechanisms and retrieval-augmented frameworks in misinformation detection systems. 

Future work should explore extending this approach to other social media platforms and further refining GNN architectures for even more nuanced content classification.

\bibliographystyle{ACM-Reference-Format}
\bibliography{sample-base}

\appendix
\section{Baseline Descriptions and Fairness of Comparisons}
In this section, we describe the baselines used for model evaluation and explain why each serves as a fair comparison for assessing the performance of our CrediRAG model with weighted GAT. The baselines span several architectures that are commonly used in fake news detection, natural language processing (NLP), and misinformation detection tasks, ensuring a robust and meaningful evaluation.

\subsection{Random Baselines}
We include two random baselines to provide a lower-bound performance comparison:
\begin{itemize}
    \item \textbf{Random:} This baseline assigns labels randomly, providing a baseline to show how much better an informed model performs compared to uninformed guessing.
    \item \textbf{Majority:} This baseline always predicts the most common class in the dataset. It serves as a fair reference for understanding the impact of class imbalance on model performance.
\end{itemize}

\subsection{Neural Network-Based Baselines}
Neural network models have been foundational in text classification and fake news detection. We include these baselines to compare against graph and transformer-based methods.
\begin{itemize}
    \item \textbf{TextCNN \cite{kim2014convolutionalneuralnetworkssentence}:} TextCNN represents a traditional neural network architecture for sentence classification. Its inclusion ensures comparison with established non-graph methods.
    \item \textbf{TextGCN \cite{yao2018graphconvolutionalnetworkstext}:} TextGCN introduces graph-based learning using word co-occurrence graphs. This baseline helps assess whether our more advanced graph neural network (GNN) approaches outperform standard graph-based text classification methods.
    \item \textbf{GAT-only \cite{velickovic2017graph}:} The GAT-only model evaluates the effectiveness of graph attention networks (GATs) without retrieval augmentation. By comparing it to retrieval-based models, we can isolate the impact of incorporating external knowledge in our model.
\end{itemize}

\subsection{Transformer-Based Baselines}
Transformers are known for achieving state-of-the-art results in various NLP tasks, including fake news detection. Thus, they serve as strong baselines for evaluating our work.
\begin{itemize}
    \item \textbf{BERT \cite{devlin2019bertpretrainingdeepbidirectional}:} BERT is a transformer model that has set benchmarks in text classification. Including it allows us to show whether our approach improves over traditional transformer architectures.
    \item \textbf{GPT3.5-turbo \cite{brown2020languagemodelsfewshotlearners}:} GPT-3.5 represents a powerful large language model with access to vast amounts of training data. Its inclusion demonstrates how our more specialized retrieval-augmented approach compares to these large-scale pretrained models.
    \item \textbf{CrediBERT \cite{Amini2024}:} CrediBERT is specifically designed for credibility assessment, making it a fair comparison against our work, which also focuses on misinformation detection.
\end{itemize}

\subsection{RAG-Based Baselines}
RAG (Retrieval-Augmented Generation) models incorporate external knowledge sources to inform their predictions. These baselines test the efficacy of retrieval-based models.
\begin{itemize}
    \item \textbf{Web-Retrieval LLM Agent \cite{tian2024web}:} This baseline retrieves real-time information from the web to enhance the language model’s factual accuracy. Testing against this model (note that the base LLM here is \textbf{GPT-4o}) ensures a fair comparison of retrieval-augmented models.
    \item \textbf{CrediRAG without GAT (Ours):} This variant of our model omits the GAT component, allowing us to measure how much graph-based reasoning contributes to the overall performance.
\end{itemize}

\subsection{GNN + RAG Baselines}
Our main contribution combines GNNs with RAG for fake news detection. These baselines show how adding graph-based attention improves retrieval-augmented methods.
\begin{itemize}
    \item \textbf{CrediRAG with un-weighted GAT (Ours):} This baseline includes an unweighted GAT in our architecture. By comparing against our weighted version, we quantify the performance gain attributed to attention-based weighting in our GNN.
\end{itemize}

\subsection{Fairness of Comparison}
It is essential to ensure that all comparisons are fair, particularly when dealing with models that may have different sources of external information. In our experiments, the AskNews knowledge base, which powers our retrieval in CrediRAG, only contains information from 2023 onward. Similarly, the Google search engine used by the Web-Retrieval LLM baseline also has access to data from 2023. Therefore, when we test on datasets like ISOT Reddit from 2017, neither retrieval system is accessing future information, ensuring that both models are evaluated under the same conditions with regard to the availability of external knowledge. This parity ensures a fair comparison between our CrediRAG model and the Web-Retrieval LLM Agent.

\section{Experiment Details}
\subsection{Dataset Statistics}
The ISOT Reddit dataset is constructed by combining Reddit submissions and comments with ground-truth labels from the ISOT Fake News dataset, which includes true and fake news articles from 2016 to 2017. The dataset is structured to match Reddit posts with news articles based on semantic similarity and publication date, creating a rich set of labeled data for misinformation detection tasks. Below are the details on how the dataset is created:

\textbf{Preprocessing:}
We begin by loading the ISOT Fake News dataset, which consists of two parts: true news articles and fake news articles. The publication dates of these articles are standardized using Python’s datetime module. Any missing or ambiguous dates are handled by using multiple date formats. The true and fake articles are then aligned to a reference date, set as January 1st, 2016, to calculate how many days have passed since that point for each article. This forms the basis for matching the news articles with Reddit posts.

\textbf{Reddit Submission and Comment Collection:}
To gather Reddit data, we process submissions and comments from relevant subreddits. The subreddits are chosen based on their relevance to political and news-related discussions during the period of interest (2016-2017). For each subreddit, we filter submissions by year, score, and number of comments to ensure that only highly engaging content is retained. Submissions are enriched with metadata, including the time of submission (in Unix time) and the number of days since the reference date.

\textbf{Matching Reddit Posts with News Articles:}
Once the Reddit submissions and comments are cleaned, we proceed to match them with articles from the ISOT Fake News dataset. Using a pre-trained SentenceTransformer model (\textit{all-MiniLM-L6-v2}), we encode the titles of both the Reddit posts and the news articles into embeddings. These embeddings are used to compute the cosine similarity between Reddit submissions and news articles within a time window of several days. This time window, set to a maximum of 3 days before and after a news article’s publication, ensures that the matched articles are temporally relevant to the Reddit discussions.

\textbf{Labeling Process:}
Submissions and comments that achieve a similarity score higher than 0.7 with fake news articles are labeled as ‘fake,’ while those matching true news articles are labeled as ‘true.’ This process is iterative and repeated for each subreddit, resulting in a comprehensive dataset of labeled Reddit posts. In addition to the matched posts, we store all associated comments to retain the contextual discussion that takes place around the news content. This ensures that our dataset captures both the post itself and the broader conversation that it generates.

\textbf{Why This Dataset is Fair for Evaluation:}
The ISOT dataset, created by matching Reddit posts to news articles in a time-constrained and context-aware manner, allows for realistic and temporally grounded misinformation detection. Importantly, the dataset does not incorporate future information, as both the fake news and Reddit discussions occurred prior to 2018. Moreover, our retrieval mechanism ensures that only relevant and semantically similar articles are matched with Reddit posts, avoiding noisy labels. The dataset serves as an ideal testbed for fake news detection models, as it replicates real-world misinformation scenarios that emerge in public discourse.
\subsection{AskNews Retrieval Process}
For retrieving news articles that are relevant to a given Reddit submission, we employ the AskNews retrieval system, which allows for searching through a large corpus of news articles while ensuring diversity in sources. However, a unique challenge arises from the constraint that AskNews only allows searches for one month of data at a time. This necessitates a month-by-month search over the entire corpus to gather the most relevant articles.

\textbf{Retrieval Methodology:}
To overcome the one-month constraint, we designed a method that iteratively searches for news articles by processing each month separately. For each Reddit submission, we execute a semantic search across all months, using natural language processing (NLP) techniques to find articles with high textual similarity to the submission. As we move through the months, we maintain a set of articles that meet the similarity threshold, ensuring that only those with a similarity score above 0.8 are considered relevant, and keeping the 50 highest similarity articles with the post content in the end.

The searching is done as follows:
\begin{verbatim}
    response = ask.news.search_news(
        query=submission_text,
        n_articles=k,
        return_type="dicts",
        method="both",               
        diversify_sources=True,    
        historical=True,           
        similarity_score_threshold=0.8
    )
\end{verbatim}

Here, the function searches for the top \(k = 20\) news articles based on their textual similarity to the Reddit submission text, while ensuring that the sources are diverse. Importantly, the retrieval is historical, so it searches through past articles rather than relying on real-time data in the past few-hours.

\textbf{Handling the Two-Month Search Constraint:}
For each submission, we call the retrieval function multiple times—once for each month in the time window of interest, using the \begin{verbatim}
    start_timestamp
\end{verbatim} and \begin{verbatim}
    end_timestamp
\end{verbatim} appropriately beginning from January 1, 2023 until present (which is the context window of AskNews). This strategy ensures that we cover all months relevant to the submission's timeline, while the high similarity threshold ensures that only highly relevant articles are retained. Additionally, the retrieval method guarantees source diversity, further ensuring that our dataset includes a wide range of perspectives, which is crucial for misinformation detection.

\textbf{Handling More Than \(k = 20\) Articles:}
Although the function is set to retrieve \(k = 20\) articles, the inclusion of neighboring articles in the graph sometimes results in retrieving more than 20 articles. This occurs because articles that are semantically related to each other may also be retrieved, as the graph structure captures the relationships between neighboring articles. This means that our final set of retrieved articles may well exceed the initial \(k = 20\), enhancing the depth and diversity of the data available for misinformation classification.

\subsubsection{Adversarial Training using GAT Model}

In our approach to classify posts as either "correct" or "misinformed," we utilize a Graph Attention Network (GAT) with adversarial training. The GAT operates on a post-to-post network, capturing relationships between posts, while adversarial perturbations help simulate noisy data and improve model robustness.

\paragraph{Graph Attention Network (GAT):}
The GAT model is designed to perform node classification by focusing on important neighboring nodes in a graph. Each node in our graph represents a post, and edges between nodes represent shared commenters. The attention mechanism in GAT allows the model to assign different importance weights to neighboring nodes during classification, helping it focus on the most relevant information. The GAT is composed of two layers: 
\begin{itemize}
    \item The first layer applies attention to the input node features (posts) and outputs hidden representations.
    \item The second layer performs binary classification (correct or misinformed) based on the attention-weighted node representations.
\end{itemize}

\paragraph{Adversarial Perturbation:}
To train the GAT model, we introduce adversarial perturbations by randomly flipping the labels of 15\% of the posts during each training epoch. This simulates real-world scenarios where some data might be mislabeled or uncertain. The GAT is trained to predict the correct labels despite these perturbations, improving its robustness to noisy or misleading data.

\paragraph{Training Process:}
The GAT model is trained using negative log likelihood loss for binary classification, and the optimizer used is Adam. Each epoch of training involves perturbing the labels, forwarding the perturbed labels through the network, and calculating the loss based on the predicted labels. This adversarial training helps the GAT model learn to identify and correct label misclassifications, improving its performance on misinformation detection tasks.

\subsubsection{Post-to-Post Graph Construction with BERT-based Stance Detection}

To build a comprehensive graph for post-to-post relationships, we leverage BERT embeddings to detect the stance between commenters and the posts they comment on. This method allows us to create weighted edges between posts, reflecting the stance of shared commenters.

\paragraph{Post-Comment Embeddings:}
We use a pre-trained BERT model from Hugging Face to generate embeddings for both posts and comments. The BERT model extracts contextualized representations of text, where each post and comment is encoded as a 768-dimensional vector. For each comment and post pair, we compute the cosine similarity between their embeddings.

\paragraph{Stance Detection using Cosine Similarity:}
Once the embeddings for both the posts and comments are generated, we calculate the cosine similarity between the embeddings to determine the stance of a comment toward a post:
\begin{itemize}
    \item If the similarity score is greater than 0.5, we classify the stance as \textbf{agree}.
    \item If the similarity score is between 0.1 and 0.5, the stance is \textbf{neutral}.
    \item If the similarity score is less than 0.1, the stance is \textbf{disagree}.
\end{itemize}
This stance detection method helps assign labels to the edges in the post-to-post graph based on how commenters interact with the posts.

\paragraph{Post-to-Post Graph Construction:}
To build the post-to-post graph, we consider pairs of posts that share commenters. If a commenter has commented on both posts, we create an edge between those two posts. The weight of the edge is determined by the stance of the shared commenter:
\begin{itemize}
    \item We calculate the stance for each shared commenter on both posts.
    \item The final edge weight is computed by multiplying the stance scores of the shared commenter on both posts and normalizing by the number of shared commenters. This allows the edge weight to reflect the overall agreement or disagreement between the posts as perceived by shared commenters.
\end{itemize}

\subsubsection{AskNews Retrieval and Ad Fontes Score Normalization}

For each Reddit submission, we retrieve relevant articles using the AskNews API. AskNews allows developers to access prompt-ready news content without the need for manual formatting. AskNews supports a range of endpoints, including news search, sentiment analysis, and Reddit context, making it versatile for chat-bots, forecasting tools, and analytical models. The retrieval process is enhanced by searching month-by-month due to the API’s constraints, ensuring that we capture historical data relevant to the submission's timeline.

\paragraph{Diversity of Sources:}
The AskNews API allows for retrieving up to 20 news articles based on textual similarity with the Reddit post. We ensure that the retrieved articles come from diverse sources by using the API’s source diversification feature. This helps prevent bias toward any particular viewpoint or outlet, making the retrieval process more fair and representative.

\paragraph{Ad Fontes Score Normalization:}
For each retrieved article, we use the Ad Fontes Media Bias Chart to obtain a reliability score for the source of the article. The Ad Fontes score ranges from 0 to 64, representing the credibility of the news source. To integrate these scores into our model, we normalize them to a range of 0 to 1 using the formula:
\[
\text{normalized\_score} = \frac{\text{Ad Fontes score}}{64}
\]
We then average the normalized scores for all retrieved articles to give an overall estimate of the reliability or misinformation level of the Reddit post. This averaged score becomes an additional feature used to classify posts as "correct" or "misinformed."
\subsubsection{Libraries and Packages Used}
To implement the entire process, we utilized several Python libraries and packages that facilitate natural language processing, graph modeling, web-based retrieval, and neural network training. Below is an overview of the core libraries and their functions:

\begin{itemize}
    \item \textbf{torch (PyTorch):} PyTorch is the primary deep learning library used for building the Graph Attention Network (GAT). It provides tensor operations, optimization functions (such as Adam), and modules for building neural network layers.
    \item \textbf{torch\_geometric:} This library extends PyTorch for handling graph-structured data, providing functionalities like graph convolution (GATConv), graph data structures, and batching mechanisms for node classification tasks.
    \item \textbf{transformers (Hugging Face):} The \texttt{transformers} library is used for loading the pre-trained BERT model and tokenizer. BERT is utilized to compute text embeddings for both posts and comments, which are then compared to detect stance.
    \item \textbf{networkx:} \texttt{networkx} is used to construct and manage the post-to-post network. It allows us to define graph structures, add edges with weights, and analyze graph properties.
    \item \textbf{scikit-learn:} The \texttt{scikit-learn} library provides tools for calculating cosine similarity between the embeddings generated by BERT. Cosine similarity is used to classify the stance between a comment and its corresponding post.
    \item \textbf{LangChain:} LangChain is a framework used to build language model agents. In our work, it is integrated to structure the google search retrieval process by leveraging agentic tools for querying relevant external sources.
    \item \textbf{DuckDuckGo Search API:} The DuckDuckGo search API is used as a part of the retrieval process to compare with the AskNews API. It helps retrieve relevant articles by performing semantic searches across the web.
    \item \textbf{AskNews API:} This external API is used to retrieve historical news articles that are semantically similar to a given Reddit post. The API also provides the ability to diversify sources to ensure balanced and fair retrieval.
    \item \textbf{Agentic Tools:} These tools are used to enable autonomous agents that retrieve and reason over external information. In particular, the LangChain agent is responsible for managing the flow of information from DuckDuckGo search results to enhance the retrieval-augmented generation (RAG) process in misinformation detection.
\end{itemize}
These libraries and tools collectively enable the construction of many robust misinformation detection pipelines that integrate graph-based reasoning, natural language processing, adversarial training, and external web-based retrieval techniques.

\end{document}